\documentclass[12pt]{article}
\usepackage{feynmf}
\usepackage{graphicx}
\usepackage{epsfig}
\newcommand{\beq}{\begin{equation}}
\newcommand{\eeq}{\end{equation}}
\newcommand{\beqn}{\begin{eqnarray}}
\newcommand{\eeqn}{\end{eqnarray}}
\begin{document} 
 
\title{About radiative kaon decay $K^+ \to \pi^+\pi^0\gamma$} 
\author{V.P. Efrosinin\\
Institute for Nuclear Research RAS\\
60th October Ave., 7A, 117312, Moscow, Russia}

\date{}
\renewcommand {\baselinestretch} {1.3}

\maketitle
\begin{abstract}

With usage of the Low theorem the general expression for amplitude of radiative
kaon decay $K^+ \to \pi^+\pi^0\gamma$ is determined. The possible reason of
suppression of a branching ratio of kaon decay $K^+ \to \pi^+\pi^0$ is
considered.    
\end{abstract}

At first, let us consider nonleptonic $K^+$ decay $K^+ \to \pi^+\pi^0$.
Suppose that this decay is suppressed by the $\Delta I=\frac{1}{2}$ rule.
It is well known that the $K^+$ is pseudoscalar $0^-$ state. Therefore,
two final pions should be in a state with zero angular momentum. This means
that full wave function of a system of two pions should be symmetrical.
Consequently, two pions can be in a state with an isospin $I=0$ or $I=2$.
In decay $K^+ \to \pi^+\pi^0$ the final state should have $I_3=1$.
Then its isospin equals to 2 with $I_3=1$. Thus, $\Delta I=\frac{3}{2}$
or $\Delta I=\frac{5}{2}$.

\begin{figure*}[hb]
\begin{center}
\epsfig{file=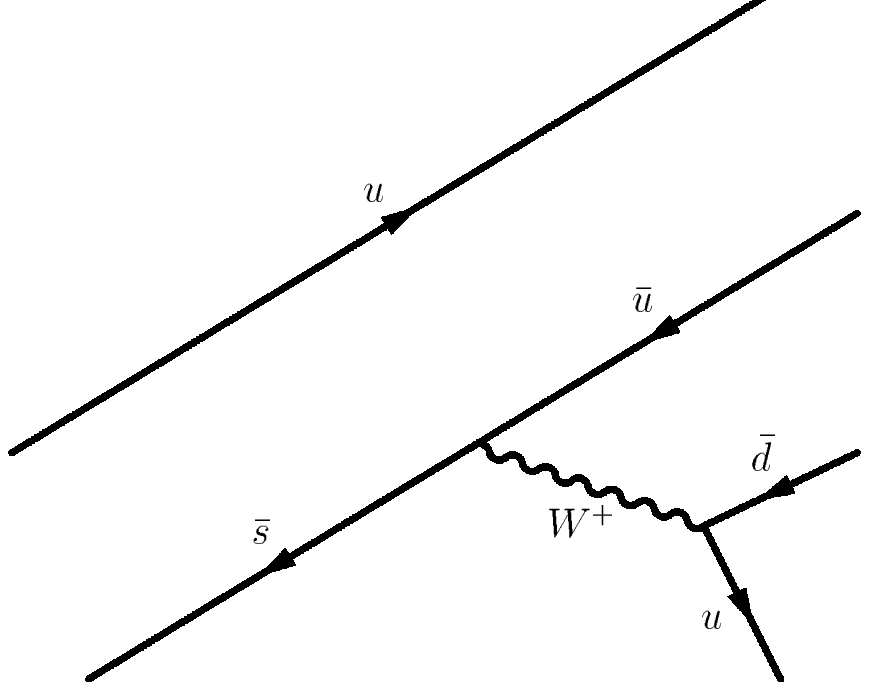,width=8.cm}
\end{center}
\caption{}
\label{fig:fig1731}
\end{figure*}

The matrix element for the process, shown by the diagram of
Fig.\ref{fig:fig1731}, 
\begin{eqnarray}
\label{eq:M1}
K^+(p) \to \pi^0(q_1)+\pi^+(q_2), 
\end{eqnarray}
where $p,q_1$ and $q_2$ denote the four-momenta of the initial $K^+$
meson, the outgoing $\pi^0$ and $\pi^+$ mesons, respectively, is given by
\begin{eqnarray}
\label{eq:M2}
T(K^+ \to \pi^0\pi^+)=\frac{G_F}{\sqrt{2}}\sin(\theta_V)\cos(\theta_A)
\frac{1}{\sqrt{2}}(p+q_1)_{\mu}\alpha f_{\pi} q^{\mu}_2.
\end{eqnarray}
Here, $G_F$ is the Fermi constant, $f_{\pi}$ is a charged pion decay
constant,
$f_{\pi}=0.131~GeV$ \cite{part}, $\theta_A$ and $\theta_V$ are the
Cabibbo angles, $\alpha$ is a dimensionless hadronisation factor, introduced
by us.
The introducing of this factor is explained by the fact that not each pair
of quark-antiquark can form a meson, but only a pair with zero total spin.
The branching ratio of the decay $K^+ \to \pi^+\pi^0$, associated with
amplitude (\ref{eq:M2}), fits experimental data $21.13 \pm 0.14\%$
\cite{part}, $21.18 \pm 0.28\%$ \cite{chiang}
at $\alpha \simeq 0.5$.

The initial and final states of radiative $K^+ \to \pi^+\pi^0\gamma$ decay
are not CP
eigenstates. Therefore, in the limit of CP conservation the amplitude of this
decay consists of the contributions: first one is an internal Bremsstrahlung,
second one is the direct photon emission of the
electrical and magnetic type. At the same
time, the measurement of the interference of internal Bremsstrahlung with
electric
dipole transitions is important for differentiating between various models of
the description of $K^+ \to \pi^+\pi^0\gamma$ decay. Therefore, in the present
article the main attention is given to the revealing of direct vector
contributions to
the amplitude of $K^+ \to \pi^+\pi^0\gamma$ decay. In our approach
we will follow
to the refs. \cite{low,chew,pest,baen,fisch}. It is known that internal
Bremsstrahlung contributions have a logarithmic singularity in $k$, that is
at integrating they behave as $k^{-1}$ for $k \to 0$. We assume that direct vector
contributions DVC and direct axial contributions DAC have no singularity for 
$k\to 0$, independently of how $k \to 0$.

\begin{figure*}[hb]
\begin{center}
\epsfig{file=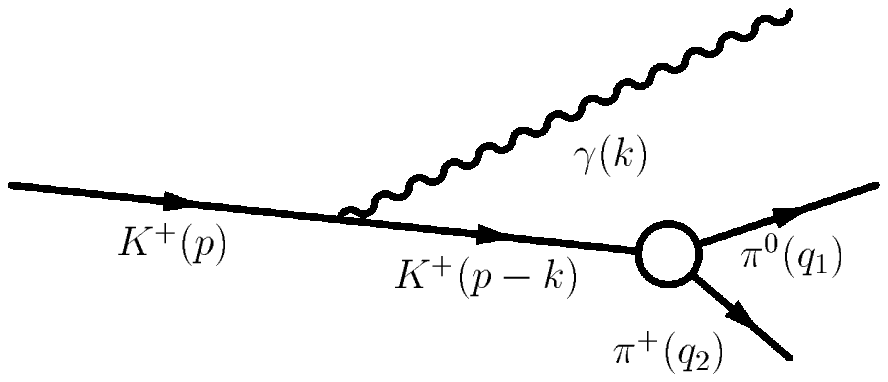,width=8.cm}
\end{center}
\caption{}
\label{fig:fig1732}
\end{figure*}

\begin{figure*}[hb]
\begin{center}
\epsfig{file=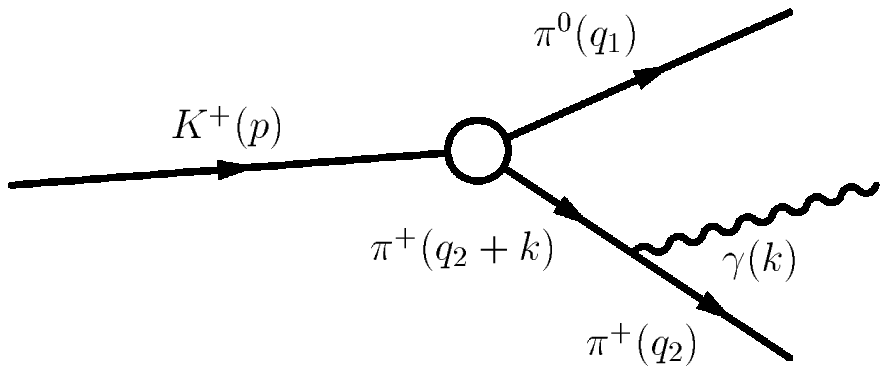,width=8.cm}
\end{center}
\caption{}
\label{fig:fig1733}
\end{figure*}

\begin{figure*}[hb]
\begin{center}
\epsfig{file=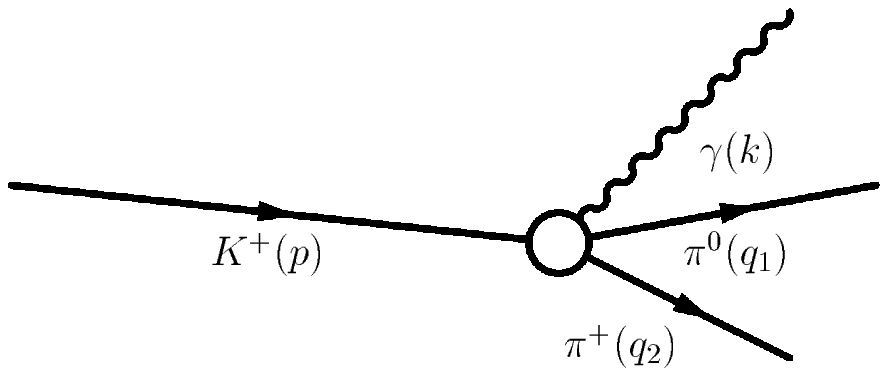,width=8.cm}
\end{center}
\caption{}
\label{fig:fig1734}
\end{figure*}

S-matrix element of kaon decay  
\begin{eqnarray}
\label{eq:M3}
K^+(p) \to \pi^0(q_1)+\pi^+(q_2)+\gamma(k) 
\end{eqnarray}
is
\begin{eqnarray}
\label{eq:M4}
<\pi^0,\pi^+,\gamma|S|K^+>=ie(2\pi)^4 \delta^{(4)}(q_1+q_2+k-p)
\varepsilon^{\mu} M_{\mu}(k),
\end{eqnarray}
where $\varepsilon^{\mu}$ is the photon polarization, $e$ is the proton
electric charge. Taking into consideration above mentioned, the quantity
$M_{\mu}(k)$ may be written as
\begin{eqnarray}
\label{eq:M5}
M_{\mu}(k)=M^{IB}_{\mu}(k)+M^{DVC}_{\mu}(k)+M^{DAC}_{\mu}(k).
\end{eqnarray}
Here, $M^{IB}_{\mu}(k)$ is the sum of terms corresponding to the photon
emission from an
external scharged lines (see Figs.\ref{fig:fig1732},\ref{fig:fig1733});
$M^{DVC}_{\mu}(k)$ and $M^{DAC}_{\mu}(k)$ are the respective sums of
direct emission
(see Fig.\ref{fig:fig1734}).
$M^{IB}_{\mu}(k)$ has the contributions of the order $k^{-1}$, $k^0$
and $O(k)$. The contributions of direct emission are of the order of $k^0$,
$k^{\ge 1}$.

From the electric current conservation follows that
\begin{eqnarray}
\label{eq:M6}
k^{\mu}M_{\mu}(k)=0.
\end{eqnarray}

Let us consider now the amplitude $T(P^2,Q^2_1,Q^2_2)$ \cite{pest}
for nonradiative off-mass shell $K^+$ decay
\begin{eqnarray}
\label{eq:M7}
K^+(P) \to \pi^0(Q_1)+\pi^+(Q_2).
\end{eqnarray}
This amplitude conserves momentum and energy but not mass. On the mass shell
the
amplitude is  $T(m^2_{K^+},m^2_{\pi^0},m^2_{\pi^+})$. For nonradiative part of
the process in Fig.\ref{fig:fig1732} the amplitude is $T((p-k)^2,m^2_{\pi^0},m^2_{\pi^+})$.
The amplitude corresponding to the process shown in the Fig.\ref{fig:fig1733}
is $T(m^2_{K^+},m^2_{\pi^0},(q_2+k)^2)$.

Accounting for these amplitudes the amplitude $M^{IB}_{\mu}(k)$ may be written
as
\begin{eqnarray}
\label{eq:M8}
M^{IB}_{\mu}(k)=\frac{(2q_2+k)_{\mu}}{(q_2+k)^2-m^2_{\pi^+}}
T(m^2_{K^+},m^2_{\pi^0},(q_2+k)^2)+T((p-k)^2,m^2_{\pi^0},m^2_{\pi^+})
\frac{(2p-k)_{\mu}}{(p-k)^2-m^2_K}.
\end{eqnarray}
Expanding $M^{IB}_{\mu}(k)$ with respect to $k$ and using the conditions
$k^2=0$, $k\cdot\varepsilon=0$, we receive
\begin{eqnarray}
\label{eq:M9}
M^{IB}_{\mu}(k)&=&\Bigl(\frac{q_{2\mu}}{q_2 \cdot k}-\frac{p_{\mu}}
{p\cdot k}\Bigr)T(m^2_{K^+},m^2_{\pi^0},m^2_{\pi^+})\nonumber\\
&+&2q_{2\mu}\frac{\partial T(m^2_{K^+},m^2_{\pi^0},Q^2_2)}
{\partial Q^2_2}\Bigg|_{Q^2_2=m^2_{\pi^+}}\nonumber\\
&+&2p_{\mu}\frac{\partial T(P^2,m^2_{\pi^0},m^2_{\pi^+})}
{\partial P^2}\Bigg|_{P^2=m^2_K}+O(k).
\end{eqnarray}
From Eqs. (\ref{eq:M5}), (\ref{eq:M6}) we obtain
\begin{eqnarray}
\label{eq:M10}
M^{DVC}_{\mu}(0)+M^{DAC}_{\mu}(0)=-2q_{2\mu}
\frac{\partial T}
{\partial Q^2_2}\Bigg|_{Q^2_2=m^2_{\pi^+}}
-2p_{\mu}\frac{\partial T}{\partial P^2}\Bigg|_{P^2=m^2_{K^+}}.
\end{eqnarray}
Then, taking into account in Eq. (\ref{eq:M5}) all terms of the order
up to $k^1$, we have
\begin{eqnarray}
\label{eq:M11}
M_{\mu}(k)&=&\Bigl(\frac{q_{2\mu}}{q_2 \cdot k}-\frac{p_{\mu}}
{p\cdot k}\Bigr)T(m^2_{K^+},m^2_{\pi^0},m^2_{\pi^+})
+O(k).
\end{eqnarray}
Let us note  that the amplitude $T(m^2_{K^+},m^2_{\pi^0},m^2_{\pi^+})$,
entering into Eq. (\ref{eq:M11}), 
is the amplitude of reaction (\ref{eq:M1}).

So, the amplitude $\varepsilon^{\mu} M_{\mu}(k)$ with accounting for also
the terms $M^{DVC}_{\mu}(k)$ and $M^{DAC}_{\mu}(k)$ in Eq. (\ref{eq:M5}) 
can be recorded as:
\begin{eqnarray}
\label{eq:M12}
\varepsilon^{\mu} M_{\mu}(k)&=&\varepsilon^{\mu}\Bigl(\frac{q_{2\mu}}
{q_2 \cdot k}-\frac{p_{\mu}}{p\cdot k}\Bigr)T(m^2_{K^+},m^2_{\pi^0},
m^2_{\pi^+})\nonumber\\
&+&\frac{G_F}{\sqrt{2}}\cos(\theta_A)\alpha f_{\pi^+}q^{\nu}_2
\Bigl(\sin({\theta}_V)<\pi^0(q_1)\gamma(k)|k^-_{\nu}|K^+(p)>\nonumber\\
&-&\sin({\theta}_A)<\pi^0(q_1)\gamma(k)|k^-_{5\nu}|K^+(p)>\Bigr).
\end{eqnarray}
Here, $<\pi^0(q_1)\gamma(k)|k^-_{\nu}|K^+(p)>$ and
$<\pi^0(q_1)\gamma(k)|k^-_{5\nu}|K^+(p)>$ are, accordingly, contributions
from $\varepsilon^{\mu} V^{DVC}_{\mu\nu}(k)$ and
$\varepsilon^{\mu} V^{DAC}_{\mu\nu}(k)$ at $k^{\geq 1}$, and
in standard identifications
\begin{eqnarray}
\label{eq:M13}
&&k^-_{\nu}=j^4_{\nu}-i j^5_{\nu},\nonumber\\
&&k^-_{5\nu}=j^4_{5\nu}-i j^5_{5\nu}.
\end{eqnarray}
The quantities $\varepsilon^{\mu} V^{DAC}_{\mu\nu}(k)$ and
$\varepsilon^{\mu} V^{DVC}_{\mu\nu}(k)$ may be covariantly decomposed as
follows:
\begin{eqnarray}
\label{eq:M14}
\varepsilon^{\mu} V^{DAC}_{\mu\nu}=\varepsilon^{\mu}\varepsilon_
{\mu\nu\alpha\beta}(bp^{\alpha}k^{\beta}+cq^{\alpha}_1k^{\beta}
+d(p \cdot k)p^{\alpha}q^{\beta}_1),
\end{eqnarray}
\begin{eqnarray}
\label{eq:M15}
\varepsilon^{\mu} V^{DVC}_{\mu\nu}&=&\varepsilon^{\mu}(A(p \cdot k)
g_{\mu\nu}
+C(p \cdot k)q_{1\mu}q_{1\nu}+D(p \cdot k)p_{\mu}p_{\nu}+Fp_{\mu}k_{\nu}
\nonumber\\
&+&G(p \cdot k)p_{\mu}q_{1\nu}+H(p \cdot k)q_{1\mu}p_{\nu}+Iq_{1\mu}k_{\nu}).
\end{eqnarray}
The coefficients in (\ref{eq:M14}), (\ref{eq:M15}) are, in general, functions
of the variables $(p \cdot k)$, $(k \cdot q_1)$.
From the structure of the contributions (\ref{eq:M14}) and (\ref{eq:M15})
it is
seen that for them the condition (\ref{eq:M6}) should be separately fulfilled:
\begin{eqnarray}
\label{eq:M16}
k^{\mu}V^{DAC}_{\mu\nu}=0,
\end{eqnarray}
\begin{eqnarray}
\label{eq:M17}
k^{\mu}V^{DVC}_{\mu\nu}=0.
\end{eqnarray}
From (\ref{eq:M16}), in view of (\ref{eq:M11}), one gets
\begin{eqnarray}
\label{eq:M18}
k^{\mu}\varepsilon_{\mu\nu\alpha\beta}(bp^{\alpha}k^{\beta}+cq^{\alpha}_1k^
{\beta}+d(p \cdot k)p^{\alpha}q^{\beta}_1)q^{\nu}_2=0.
\end{eqnarray}
From here it follows that
\begin{eqnarray}
\label{eq:M19}
d=0,
\end{eqnarray}
and
\begin{eqnarray}
\label{eq:M20}
b\varepsilon_{\mu\nu\alpha\beta}p^{\alpha}k^{\beta}q^{\nu}_2=
b\varepsilon_{\mu\nu\alpha\beta}q^{\alpha}_1k^{\beta}q^{\nu}_2.
\end{eqnarray}
Therefore, there is only one contribution tu the $V^{DAC}_{\mu\nu}$:
\begin{eqnarray}
\label{eq:M21}
V^{DAC}_{\mu\nu}=b^{\prime}\varepsilon_{\mu\nu\alpha\beta}q^{\alpha}_1k^{\beta}.
\end{eqnarray}
Further, in view of Eq. (\ref{eq:M15}), one obtains
\begin{eqnarray}
\label{eq:M22}
k^{\mu}V^{DVC}_{\mu\nu}&=&(A(p \cdot k)+F(p \cdot k)+I(q_1 \cdot k))k_{\nu}
\nonumber\\
&+&(D(p \cdot k)(p \cdot k)+H(p \cdot k)(q_1 \cdot k))p_{\nu}\nonumber\\
 &+&(C(p \cdot k)(q_1 \cdot k)+G(p \cdot k)(p \cdot k))q_{1\nu}=0.
\end{eqnarray}
From here we receive the following relations among the coefficients,
entering into Eq. (\ref{eq:M22}),
\begin{eqnarray}
\label{eq:M23}
F&=&-A-I\frac{(q_1 \cdot k)}{(p \cdot k)},\nonumber\\
D&=&-H\frac{(q_1 \cdot k)}{(p \cdot k)},\nonumber\\
G&=&-C\frac{(q_1 \cdot k)}{(p \cdot k)}.
\end{eqnarray}

Using these relations, we have
\begin{eqnarray}
\label{eq:M24}
V^{DVC}_{\mu\nu}&=&A((p \cdot k)g_{\mu\nu}-p_{\mu}k_{\nu})\nonumber\\
&+&C((p \cdot k)q_{1\mu}q_{1\nu}-(q_1 \cdot k)p_{\mu}q_{1\nu})\nonumber\\
&+&H((p \cdot k)q_{1\mu}p_{\nu}-(q_1 \cdot k)p_{\mu}p_{\nu})\nonumber\\
&+&I(q_{1\mu}k_{\nu}-p_{\mu}k_{\nu}\frac{(q_1 \cdot k)}{(p \cdot k)}).
\end{eqnarray}
Finally, the amplitude $\varepsilon^{\mu}M_{\mu}(k)$ becomes
\begin{eqnarray}
\label{eq:M25}
\varepsilon^{\mu}M_{\mu}(k)&=&\varepsilon^{\mu}
\Bigl(\frac{q_{2\mu}}{q_2 \cdot k}-\frac{p_{\mu}}
{p\cdot k}\Bigr)T(m^2_{K^+},m^2_{\pi^0},m^2_{\pi^+})\nonumber\\
&+&\frac{G_F}{\sqrt{2}}\cos(\theta_A)\alpha f_{\pi^+}\varepsilon^{\mu}
q^{\nu}_2\Bigl(\sin(\theta_V)V^{DVC}_{\mu\nu}-\sin(\theta_A)V^{DAC}_{\mu\nu}
\Bigr).
\end{eqnarray}
The quantities $V^{DVC}_{\mu\nu}$ and $V^{DAC}_{\mu\nu}$, entering into
Eq. (\ref{eq:M25}),
are expressed by the formulas
(\ref{eq:M24}) and (\ref{eq:M21}), accordingly.

In our calculations we assume that the contributions of direct radiation
are insignificant.
This is confirmed by results of KEK-E470 experiment on measurement of the direct
photon emission in $K^+ \to \pi^+\pi^0\gamma$ decay \cite{aliev}.
Such, here the best
fit to the $\pi^+$ decay spectrum in the range of $\pi^+$
kinetic energy in c.m.s. from 55 to 90 $MeV$ is obtained for the
braching ratio for the direct photon emission
(magnetic dipole M1 transition) of
$[3.8\pm0.8(stat)\pm0.7(syst)]\times 10^{-6}$.
From the previous
measurements of the direct photon emission in $K^+ \to \pi^+\pi^0\gamma$ decay
follows that this ratio is
$[4.7 \pm 0.8 \pm 0.3]\times 10^{-6}$ \cite{adler}, and in
$K^- \to \pi^-\pi^0\gamma$ decay it equals to
$[3.7 \pm 3.9 \pm 1.0]\times 10^{-6}$ \cite{uvarov}. The theoretical prediction
for the branching ratio of the DE component is the following:
$Br(DE)=3.5 \times 10^{-6}$
\cite{ecker1,bijnens}.

Then, using Eq. (\ref{eq:M4}),
the cross section of $K^+ \to \pi^+\pi^0\gamma$ decay is determined with
accounting for only
the
main contributions from the Bremsstrahlung term (\ref{eq:M11})
excluding $O(k)$.
Adopting it,
we receive a branching ratio of $K^+ \to \pi^+\pi^0\gamma$ decay $2.84\times
10^{-4}$ in the $\pi^+$ kinetic region from 55 to 90 $MeV$.
The obtained ratio does not
contradict of
experimental values of $(2.75 \pm 0.15)\times 10^{-4}$ \cite{part},
$(2.71 \pm 0.45)\times 10^{-4}$ \cite{bolotov},
$(2.87 \pm 0.32)\times 10^{-4}$ \cite{smith},
$(2.71 \pm 0.19)\times 10^{-4}$ \cite{abrams}.

Nevertheless, determination from experiment of an interference of the
direct electrical
contributions of the amplitude (\ref{eq:M25}) and Bremsstrahlung can 
check the chiral perturbation theory \cite{ecker1,ecker2}.

In summary, in the present paper we have obtained the general expression for
the amplitude of radiative kaon decay $K^+ \to \pi^+\pi^0\gamma$. It is shown
that the Bremsstrahlung terms are dominant in this decay.





\newpage
\clearpage

\newpage
\clearpage


\newpage
\clearpage

\newpage
\clearpage

\newpage
\clearpage

\begin{center}
Figure captions
\end{center}

Fig.~1. Nonradiative decay $K^+ \to \pi^0+\pi^+$.
		
Fig.~2. Diagram give the $K^+$ inner Bremstrahlung term of
$K^+ \to \pi^++\pi^0+\gamma$ decay.			
		
Fig.~3. Diagram give the $\pi^+$ inner Bremstrahlung term of
$K^+ \to \pi^++\pi^0+\gamma$ decay. 

Fig.~4. Direct vector and axial contributions to the
$K^+ \to \pi^++\pi^0+\gamma$  decay.


\end{document}